\documentstyle [12pt]{article}
\advance\textwidth by +1.0in
\advance\textheight by +1.0in
\advance\oddsidemargin by -0.5in
\advance\evensidemargin by -1.0in
\advance\topmargin by -0.5in
\def\ov#1{\overline{#1}}

\def\vb#1{\mbox{\boldmath$#1$}}
\def\pd#1#2{\frac{\partial #1}{\partial #2}}

\def\bdot{\,\vb{\cdot}\,}
\def\btimes{\,\vb{\times}\,}

\begin{document}

\begin{flushright}
October 10, 2001
\end{flushright}

\begin{center}

{\sf A Geometric View of Hamiltonian Perturbation Theory} \\

\vspace*{0.2in}

Alain J.~Brizard\footnote{Phone: (802) 654-2886; Fax: (802) 654-2236; Email: abrizard@smcvt.edu} \\
{\it Department of Chemistry and Physics, Saint Michael's College} \\
{\it One Winooski Park, Colchester, Vermont 05439}

\end{center}

\vspace*{0.2in}

The variational formulation for Lie-transform Hamiltonian perturbation theory is presented in terms of an action functional defined on a two-dimensional parameter space. A fundamental equation in Hamiltonian perturbation theory is shown to result from the freedom of choice of 
the integration path for the action functional.

\vspace*{0.4in}

\noindent
Keywords: Lie transform, Hamiltonian perturbation theory

\vspace*{0.2in}

\noindent
PACS Numbers: 03.20.+i, 52.20.Dq

\vfill\eject

The paradigm of canonical Hamiltonian perturbation theory \cite{RGL_1,LL} involves the transformation of an {\it exact} Hamiltonian $H_{\epsilon}$, which depends continuously on a perturbation parameter $\epsilon$, into a {\it reference} Hamiltonian $H_{0}$ for which the Hamilton equations $\partial_{t}{\bf z} = \{ {\bf z},\; H_{0} \}$ have a known solution (unless otherwise noted ${\bf z}$ denotes canonical phase-space coordinates and $\{\;,\;\}$ denotes the canonical Poisson bracket). According to the Lie-transform approach to Hamiltonian perturbation theory \cite{RGL_1}, the transformation $H_{\epsilon} \rightarrow H_{0}$ is {\it induced} by a reversible (time-dependent) phase-space transformation from the old phase-space coordinates ${\bf z}$ to the new phase-space coordinates $\ov{{\bf z}}(\epsilon) = T_{\epsilon}\;{\bf z}$, where $T_{\epsilon}$ is an operator defined in terms of a generating scalar field $S_{\epsilon}$ as
\[ T_{\epsilon} \;\equiv\; \exp \left( \int_{0}^{\epsilon}\; \{ S_{\sigma},\; \;\} \;d\sigma \right). \]
For a time-dependent phase-space transformation ${\bf z} \rightarrow \ov{{\bf z}}(\epsilon)$ generated by the scalar field $S_{\epsilon}({\bf z},t)$, the transformation from the old Hamiltonian $H_{\epsilon}$ to the new Hamiltonian $H_{0}$ is expressed as \cite{LL,Goldstein}
\begin{equation} 
H_{0}(\ov{{\bf z}}(\epsilon),t) \;\equiv\; H_{\epsilon}({\bf z},t) \;-\; \int_{0}^{\epsilon}\;
\pd{S_{\sigma}({\bf z},t)}{t}\;d\sigma. 
\label{eq:H_epsilon}
\end{equation}
The transformations ${\bf z} \rightarrow \ov{{\bf z}}$ and $H_{\epsilon} \rightarrow H_{0}$ are therefore completely determined by the perturbed Hamiltonian $H_{\epsilon}({\bf z},t)$ and the phase-space generating function $S_{\epsilon}(
{\bf z},t)$. Since $H_{0}$ is independent of $\epsilon$ by construction (i.e., $\partial_{\epsilon}H_{0} \equiv 0$), the $\epsilon$-derivative of both sides in (\ref{eq:H_epsilon}) yields a dynamical evolution (henceforth known as the Lie-transform perturbation equation) 
\begin{equation}
\partial_{t}S \;+\; \{ S,\; H \} \;=\; \partial_{\epsilon}H,
\label{eq:Lie}
\end{equation} 
where the parametric $\epsilon$-dependence is included with the time dependence. The $\epsilon$-perturbed 
Hamilton equations, on the other hand, are now expressed as
\begin{equation}
\pd{z^{\alpha}(t,\epsilon)}{t} \;=\; \left\{ z^{\alpha},\; H({\bf z};t,\epsilon) \right\},
\label{eq:Ham_t}
\end{equation}
and
\begin{equation}
\pd{z^{\alpha}(t,\epsilon)}{\epsilon} \;=\; \left\{ z^{\alpha},\; S({\bf z};t,\epsilon) \right\}.
\label{eq:Ham_epsilon}
\end{equation}
Hence, whereas the Hamiltonian $H$ is the generating function for the infinitesimal canonical transformations associated with the dynamical evolution (\ref{eq:Ham_t}) of the Hamiltonian system (henceforth referred to as the $t$-dynamics), the generating function $S$ is the Hamiltonian for the {\it perturbation} evolution (\ref{eq:Ham_epsilon}) of the Hamiltonian system (henceforth referred to as the $\epsilon$-dynamics). We note that the order with which the Hamiltonian system is evolved and perturbed should be immaterial, i.e., the same dynamical state ${\bf z}(t,\epsilon) \equiv 
{\bf z}_{\epsilon}(t)$ can be reached by either evolving the unperturbed system first $[{\bf z}_{0}(t = 0) \rightarrow 
{\bf z}_{0}(t)]$ and then perturbing it $[{\bf z}_{0}(t) \rightarrow {\bf z}_{\epsilon}(t)]$ or perturbing the system first $[{\bf z}_{0}(t = 0) \rightarrow {\bf z}_{\epsilon}(t = 0)]$ and then evolving it $[{\bf z}_{\epsilon}(t = 0) \rightarrow 
{\bf z}_{\epsilon}(t)]$.

The purpose of this Letter is to present the variational formulation for the Lie-transform perturbation equation (\ref{eq:Lie}) and the multi-Hamilton equations (\ref{eq:Ham_t}) and
(\ref{eq:Ham_epsilon}). In particular, we show that the Lie-transform perturbation equation 
(\ref{eq:Lie}) is a direct consequence of path independence in our variational formulation. 
First, we introduce an extended phase-space Lagrangian $\Gamma$ on the $(t,\epsilon)$-plane defined as
\begin{equation}
\Gamma({\bf Z}(t,\epsilon);t,\epsilon) \;\equiv\; {\bf P}(t,\epsilon)\;\vb{\cdot}\; d{\bf Q} (t,\epsilon) \;-\; 
H({\bf Z}(t,\epsilon);t,\epsilon)\; dt \;-\; S({\bf Z}(t,\epsilon);
t,\epsilon)\; d\epsilon,
\label{eq:Gamma}
\end{equation}
where
\begin{equation}
Z^{\alpha}: \;\;(t,\epsilon) \;\mapsto\; Z^{\alpha}(t,\epsilon) \equiv ({\bf Q}(t,\epsilon),
{\bf P}(t,\epsilon)) 
\label{eq:mapping}
\end{equation}
denotes a generic mapping from the $(t,\epsilon)$-plane to the $2N$-dimensional phase space, with
\[ dZ^{\alpha}(t,\epsilon) \;\equiv\; \pd{Z^{\alpha}(t,\epsilon)}{t} \;dt \;+\;
\pd{Z^{\alpha}(t,\epsilon)}{\epsilon} \;d\epsilon. \]
(In what follows, the uppercase ${\bf Z}$ denotes a generic mapping whereas the lowercase ${\bf z}$ denotes a Hamiltonian
orbit in phase space.)
The one-form (\ref{eq:Gamma}) is said to be extended in the sense that the $\epsilon$-dynamics Hamiltonian term $-\,S\,d\epsilon$ has been added to the standard phase-space Lagrangian 
${\bf P}\bdot d{\bf Q} - H\,dt$. Next, we define the action integral
\begin{equation}
{\cal A}_{C}[{\bf Z}] \;\equiv\; \int_{C}\;\Gamma({\bf Z}(t,\epsilon);t,\epsilon),
\label{eq:action_C}
\end{equation}
where $C$ denotes an arbitrary path between two (distinct) points on the two-dimensional $(t,\epsilon)$-plane. For a fixed path $C$, the action integral ${\cal A}_{C}[{\bf Z}]$ is a functional of the mapping (\ref{eq:mapping}). Holding $C$ fixed, we first consider the variational principle $\delta{\cal A}_{C}[{\bf z}] = 0$ corresponding to an arbitrary variation $\delta{\bf Z} \equiv {\bf Z} - {\bf z}$ (which is assumed to vanish at the end points of $C$). Using Eqs.~(\ref{eq:Gamma}) and (\ref{eq:action_C}), we thus find
\begin{equation}
\delta{\cal A}_{C}[{\bf z}] \;=\; \int_{C}\; \delta Z^{\alpha} \left(\;\omega_{\alpha\beta}\; dz^{\beta} \;-\; \pd{H}{z^{\alpha}}\;dt \;-\; \pd{S}{z^{\alpha}} \;d\epsilon \;\right) \;+\; \int_{C}\; d\left( {\bf p}\; \vb{\cdot}\;\delta{\bf Q} \right),
\label{eq:delta_A}
\end{equation}
where the second path integral vanishes since we assumed that $\delta{\bf Z}$ vanishes at the end points of $C$; here, $\omega_{\alpha\beta}$ denotes the components of the canonical Lagrange tensor (i.e., $dp_{i}\wedge dq^{i} = \frac{1}{2}\,\omega_{\alpha\beta}\;dz^{\alpha}\wedge dz^{\beta}$). The multi-Hamilton equations (\ref{eq:Ham_t}) and (\ref{eq:Ham_epsilon}) are automatically recovered in Euler-Lagrange form as $\omega_{\alpha\beta}\;dz^{\beta} \;=\; \partial_{\alpha}H\,dt + \partial_{\alpha}S\,d\epsilon$ from the variational principle $\delta {\cal A}_{C}[{\bf z}] = 0$ for arbitrary variations $\delta{\bf Z}$ and an arbitrary path $C$.

In the standard variational principle for single-particle Hamiltonian dynamics \cite{Goldstein},
the action functional involves a time integration of the phase-space Lagrangian ${\bf P}\bdot d{\bf Q} - H\,dt$ from an initial time $t_{1}$ to a final time $t_{2}$; the initial and final times play no role and the variational principle yields the standard Hamilton equations. The variational principle based on the action functional (\ref{eq:action_C}) presents us with a new problem: determining how the freedom of choice in selecting the path $C$ on the $(t,\epsilon)$-plane is expressed mathematically. To resolve this problem, we first consider two different paths $C$ and $C'$, both having the same end points, and we choose the mapping (\ref{eq:mapping}) to be a multi-Hamiltonian orbit ${\bf z}(t,\epsilon)$, i.e., a solution of Eqs.~(\ref{eq:Ham_t}) and (\ref{eq:Ham_epsilon}). Next, according to Stokes' theorem \cite{Spivak}, the difference between ${\cal A}_{C}[{\bf z}]$ and ${\cal A}_{C'}[{\bf z}]$ is evaluated as
\begin{equation}
{\cal A}_{C}[{\bf z}] \;-\; {\cal A}_{C'}[{\bf z}] \;=\; \oint_{C - C'}\;\Gamma \;=\; \int_{{\cal D}}\; d\Gamma,
\label{eq:Poincare}
\end{equation}
where ${\cal D}$ denotes the area in the $(t,\epsilon)$-plane enclosed by the two paths $C$ and $C'$ (i.e., $C - C' \equiv \partial{\cal D}$ denotes the contour of ${\cal D}$). Using Eqs.~(\ref{eq:Ham_t})-(\ref{eq:Gamma}), the two-form $d\Gamma$ appearing in
Eq.~(\ref{eq:Poincare}) is
\begin{equation}
d\Gamma(t,\epsilon;{\bf z}(t,\epsilon)) \;=\; d\epsilon\wedge dt \left(\; \frac{dS}{dt} \;-\; \frac{dH}{d\epsilon} \;-\; \{ S,\; H\} \;\right),
\label{eq:dGamma}
\end{equation}
where the operators $d/dt$ and $d/d\epsilon$ are defined as
\begin{equation}
\left. \begin{array}{rcl}
d/dt & \equiv & \partial_{t} \;+\; \{ \;,\; H \} \\
 &  & \\
d/d\epsilon & \equiv & \partial_{\epsilon} \;+\; \{\;,\; S \}
\end{array} \right\}.
\label{eq:dt_depsilon}
\end{equation}
From Eq.~(\ref{eq:Poincare}), we see that, for a fixed Hamiltonian orbit ${\bf z}(t,\epsilon)$,
the condition for the path-independence of the action integral ${\cal A}_{C}[{\bf z}]$ (i.e., ${\cal A}_{C}[{\bf z}] = {\cal A}_{C'}[{\bf z}]$) is $d\Gamma = 0$. From Eq.~(\ref{eq:dGamma}), we see that this condition holds provided $H$ and $S$ satisfy the following constraint equation:
\begin{equation}
\pd{S}{t} \;-\; \pd{H}{\epsilon} \;=\; \{ H,\; S \},
\label{eq:constraint}
\end{equation}
which is exactly the Lie-transform perturbation equation (\ref{eq:Lie}) appearing in Lie-transform Hamiltonian perturbation theory. We can also verify by using the operators defined in (\ref{eq:dt_depsilon}) and the Jacobi identity for the Poisson bracket that the constraint (\ref{eq:constraint}) ensures that the flows associated with $t$-dynamics and 
$\epsilon$-dynamics commute:
\begin{equation}
\left[ \frac{d}{dt},\; \frac{d}{d\epsilon} \right]\;g({\bf z}; t,\epsilon) \;=\;
\frac{d}{dt} \left( \frac{dg}{d\epsilon} \right) \;-\; \frac{d}{d\epsilon} \left(
\frac{dg}{dt} \right) \;=\; 0
\label{eq:commute}
\end{equation}
for all $g({\bf z}; t,\epsilon)$. As expected the interpretation of the commutation relation (\ref{eq:commute}) is that the order with which the Hamiltonian system is evolved (under the $t$-dynamics) and is perturbed (under the $\epsilon$-dynamics) is indeed immaterial.
 
So far we have presented the variational formulation of Lie-transform canonical Hamiltonian perturbation theory. Most 
recent applications of Hamiltonian perturbation theory in plasma physics are carried out using non-canonical phase-space coordinates \cite{RGL_2,AJB}. Non-canonical Lie-transform Hamiltonian perturbation theory possesses a variational formulation expressed in terms of the extended non-canonical phase-space Lagrangian 
\begin{equation} 
\Gamma \;=\; \left[\; m{\bf v} \;+\; \frac{e}{c}\;{\bf A}({\bf x};t,\epsilon) \;\right]\bdot d{\bf x} \;-\; H({\bf x},
{\bf v};t,\epsilon)\;dt \;-\; S({\bf x},{\bf v};t,\epsilon)\;d\epsilon,
\label{eq:Lie_nc}
\end{equation}
where the non-canonical coordinates $({\bf x},{\bf v})$ denote the particle position and its velocity while ${\bf A}(
{\bf x};t,\epsilon)$ denotes the perturbed magnetic vector potential. For a fixed path $C$ in the $(t,\epsilon)$-plane, 
the variational principle $\delta(\int_{C}\Gamma) = 0$ yields the Euler-Lagrange equations
\begin{eqnarray*}
m\,d{\bf x} & = & \pd{H}{{\bf v}}\;dt \;+\; \pd{S}{{\bf v}}\;d\epsilon \\
 &   & \\
m\,d{\bf v} & = & - \left( \nabla H \;+\; \frac{e}{c}\,\pd{{\bf A}}{t} \right) dt \;-\; 
\left( \nabla S \;+\; \frac{e}{c}\,\pd{{\bf A}}{\epsilon} \right) d\epsilon \;+\; \frac{e}{c}\;d{\bf x}\btimes{\bf B}, 
\end{eqnarray*}
which can be written in multi-Hamilton form as
\begin{equation} 
\left. \begin{array}{rcl}
\partial_{t}z^{\alpha} & = & \{ z^{\alpha},\; H\}_{{\rm nc}} \;+\; (e/c)\;\partial_{t}
{\bf A}\bdot\{ {\bf x},\; z^{\alpha} \}_{{\rm nc}} \\
 &  & \\
\partial_{\epsilon}z^{\alpha} & = & \{ z^{\alpha},\; S \}_{{\rm nc}} \;+\; (e/c)\;\partial_{\epsilon}{\bf A}\bdot\{ 
{\bf x},\; z^{\alpha} \}_{{\rm nc}}
\end{array} \right\},
\label{eq:LTPE}
\end{equation}
where $\{\;,\;\}_{{\rm nc}}$ denotes the non-canonical Poisson bracket
\[ \{ a,\; b\}_{{\rm nc}} \;\equiv\; \frac{1}{m} \left( \nabla a\bdot\pd{b}{{\bf v}} \;-\; \pd{a}{{\bf v}}\bdot\nabla b \right) \;+\; \frac{e{\bf B}}{m^{2}c}\bdot \left( \pd{a}{{\bf v}}
\btimes\pd{b}{{\bf v}} \right). \]
The condition $d\Gamma = 0$ for path independence, on the other hand, yields the Lie-transform non-canonical perturbation equation
\begin{equation}
\pd{S}{t} \;-\; \pd{H}{\epsilon} \;=\; \{ H,\; S\}_{{\rm nc}} \;+\; \frac{e}{mc} \left(
\pd{{\bf A}}{t}\bdot\pd{S}{{\bf v}} \;-\; \pd{{\bf A}}{\epsilon}\bdot\pd{H}{{\bf v}} \right).
\label{eq:Lie_nc}
\end{equation}
This non-canonical Hamiltonian perturbation equation plays a prominent role in the derivation of the reduced nonlinear gyrokinetic equations describing the perturbed Hamiltonian dynamics of charged particles under the influence of 
low-frequency electromagnetic fluctuations in a magnetized plasma \cite{AJB}. A linearized version of (\ref{eq:Lie_nc}) 
also plays a prominent role in the free-energy method developed by Morrison and Pfirsch \cite{MP_89,MP_90,PM_91} to investigate the linear stability of various plasma equilibria.

Finally, we note that the variational formulation for Lie-transform Hamiltonian perturbation theory can be generalized to include several perturbation parameters. Consider the $(k + 1)$-component perturbation vector $\vb{\epsilon} \equiv (\epsilon^{0},\epsilon^{1},...,\epsilon^{k})$, where each perturbation parameter $\epsilon^{a}$ is paired with a Hamiltonian $S_{a}$ (with $\epsilon^{0} \equiv t$ and $S_{0} \equiv H$). Hence, the multi-Hamilton equations (\ref{eq:Ham_t}) and (\ref{eq:Ham_epsilon}) are now replaced by the set of multi-Hamilton equations:
\begin{equation}
\pd{z^{\alpha}(\vb{\epsilon})}{\epsilon^{a}} \;=\; \left\{ z^{\alpha},\; S_{a}({\bf z};\vb{\epsilon}) \right\},
\label{eq:Ham_multi}
\end{equation}
where $a = 0,1,...,k$. We now define the action functional ${\cal A}_{C}[{\bf Z}] \equiv \int_{C} \Gamma({\bf Z}(
\vb{\epsilon});\vb{\epsilon})$ in terms of a path $C$ in the $(k+1)$-dimensional $\vb{\epsilon}$-space and the one-form:
\begin{equation}
\Gamma({\bf Z}(\vb{\epsilon});\vb{\epsilon}) \;\equiv\; {\bf P}(\vb{\epsilon})\;\vb{\cdot}\; 
d{\bf Q}(\vb{\epsilon}) \;-\; S_{a}({\bf Z}(\vb{\epsilon});\vb{\epsilon})\; d\epsilon^{a},
\label{eq:Gamma_multi}
\end{equation}
where ${\bf Z}: \vb{\epsilon} \mapsto {\bf Z}(\vb{\epsilon})$ denotes a generic mapping from $\vb{\epsilon}$-space to phase space. For a fixed (but arbitrary) path $C$, we recover the multi-Hamilton equations (\ref{eq:Ham_multi}) from the variational principle $\delta\,{\cal A}_{C}[{\bf z}] = 0$. In addition, for a fixed multi-Hamiltonian orbit ${\bf z}(
\vb{\epsilon}) = ({\bf q}(\vb{\epsilon}),\; {\bf p}(\vb{\epsilon})$, the $(k+1)$ equations corresponding to the condition for the path-independence of ${\cal A}_{C}[{\bf z}]$ (i.e., $d\Gamma = 0$) are
\begin{equation}
\partial_{a}S_{b} \;-\; \partial_{b}S_{a} \;=\; \{ S_{a},\; S_{b} \}.
\label{eq:constraint_multi}
\end{equation}
Note that this set of equations is invariant under the gauge transformation $S_{a} \rightarrow S_{a} - \partial\chi/
\partial\epsilon^{a}$, where $\chi({\bf q};\vb{\epsilon})$ is an arbitrary scalar field. Gauge invariance
of the multi-Hamiltonian dynamics requires that the phase-space Lagrangian $\Gamma$ transform as $\Gamma \rightarrow
\Gamma + d\chi$ and ${\bf p} \rightarrow {\bf p} + \partial\chi/\partial{\bf q}$. The gauge properties of the generating function $S$ are further elucidated in Ref.~\cite{AJB_94}.

The work presented here introduced a variational formulation for Lie-transform Hamiltonian perturbation theory. This work follows recent developments in the variational formulation of exact and reduced Vlasov-Maxwell equations \cite{Larsson_91,AJB_94,AJB_prl,AJB_00} as well as other important dynamical field equations in plasma physics (e.g., kinetic-MHD equations \cite{AJB_94} and drift-wave equations \cite{drift}).

{\it This work was supported by the U.~S.~Department of Energy under Contract No.~DE-AC03-76SFOO098.}

\vfill\eject

\end{document}